\documentclass[12pt]{iopart}

\usepackage{graphicx}
\usepackage{iopams}  
\usepackage{subcaption}
\usepackage[ruled,linesnumbered,lined, noend]{algorithm2e}
\SetKwProg{Fn}{}{:}{}
\usepackage[braket]{qcircuit}
\usepackage{hyperref}
\usepackage{xcolor} 
\newcommand{\data}{data} % dataset
\newcommand{\D}{\mathcal {D}}

\begin{document}

\title[Double sparse quantum state preparation]{Double sparse quantum state preparation}

\author{\href{https://orcid.org/0000-0003-1840-4276}{Tiago M.L. de Veras$^{1,2}$}, \href{https://orcid.org/0000-0002-2823-3418}{Leon D. da Silva$^2$} and \href{https://orcid.org/0000-0003-0019-7694}{Adenilton J. da Silva}$^1$}

\address{$^1$Centro de Inform\'{a}tica, Universidade Federal de Pernambuco, 50670-901 Recife, PE, Brazil}
\address{$^2$Departamento de Matem\'atica, Universidade Federal Rural de Pernambuco, 52171-900 Recife, PE, Brazil}

\ead{tiago.veras@ufrpe.br}
\vspace{10pt}
\begin{indented}
\item[]
\end{indented}

\begin{abstract}
Initializing classical data in a quantum device is an essential step in many quantum algorithms. As a consequence of measurement and noisy operations, some algorithms need to reinitialize the prepared state several times during its execution.  In this work, we propose a quantum state preparation algorithm called CVO-QRAM with computational cost $\mathcal{O}(kM)$, where $M$ is the number of nonzero probability amplitudes and $k$ is the maximum number of bits with value 1 in the patterns to be stored. The proposed algorithm can be an alternative to create sparse states in future NISQ devices.

\end{abstract}

\vspace{2pc}

\section{Introduction}

Transferring information from a classical computer to a quantum device can negate the advantages obtained in quantum computing, as the cost of transferring the information to the quantum device can dominate the cost of the algorithm~\cite{tang2021quantum}. Information transfer from a classical device to a quantum device presents difficulties that are not found in the transfer of information between two classical devices. With classical devices, one can transfer the information to an auxiliary device only once and carry out all the necessary processing. Due to the decoherence of information, the impossibility of copying quantum states, and the noisy execution of algorithms, it may be necessary to transfer information to the quantum devices repeatedly. Given the classical data in Eq.~\ref{eq:data}, 
\begin{equation}
\label{eq:data}
    \mathcal {D} = \left \lbrace(x_k, p_k) | x_k \in \mathbb{C}, \sum_{k=0}^{M-1} | x_k |^2 = 1, p_k \in \lbrace 0,1 \rbrace^{n} \right \rbrace,
\end{equation}
a quantum state preparation algorithm is a classical algorithm that creates a circuit $SP_\mathcal{D}$, where $SP_\mathcal{D}\ket{0}$ is equal to the state $\ket{\psi}$ described in Eq.~\ref{eq:amp_enc}.
\begin{equation}
\label{eq:amp_enc}
     |\psi \rangle = \sum_{k = 0}^{M-1} x_k | p_k \rangle.
\end{equation}

One can classify state preparation algorithms as exact algorithms~\cite{ventura1999initializing,Long_2001,Shende2006,Park2019,veras2020}, and approximated algorithms~\cite{grover2000synthesis,Soklakov2006,Sanders2019,bausch2020fast}. This work focus on the exact state preparation algorithms. The exact state preparation can be grouped into two types: i)  algorithms that prepare the quantum states, loading each pattern in a quantum superposition one by one, with computational cost related to the number of amplitudes and qubits~\cite{ventura1999initializing,Park2019,veras2020}; ii) algorithms that use decompositions of quantum states to prepare the state with exponential computational cost in relation to the number of qubits of the desired state \cite{mottonen2005transformation,Shende2006,plesch2011quantum}. Algorithms with exponential cost concerning the number of qubits and input patterns are not efficient, and can only be used to generate quantum states with a small number of qubits. The algorithms with computational cost $O(nM)$ require a high number of CNOTs and are not suitable for NISQ devices.

This paper aims to develop an algorithm to transfer sparse data to quantum devices with computational cost $O(M log(M)+nM)$ for the construction of the quantum circuit by classical computers and that produces quantum circuits with a small number of CNOT operators when compared with previous algorithms in the literature. To achieve this goal we optimize the Continuous-Valued QRAM~\cite{veras2020} defining a partial order of presentation of the data in $\D$. When compared with the sparse quantum state preparation algorithm recently proposed in~\cite{sparse_isometries_2021} whose computational cost is $O(M^2+nM)$ for the construction of the quantum circuit by classical computers, our method generates circuits with a lower number of CNOT gates in the double sparse case (sparse concerning amplitudes and the number of 1s in the state).

The rest of this work is divided in 5 parts. Section \ref{sec:quantumoperators} describes the quantum operators used in this work. Section~\ref{sec:CV-QRAM algorithm} presents the CV-QRAM algorithm~\cite{veras2020}.  Section \ref{sec:CVO-QRAM} describes the CVO-QRAM algorithm proposed in this work. Section \ref{sec:Results_and_experiments} presents the results of the experiments performed and shows the improvements achieved by the proposed algorithm. Finally,  Section \ref{sec:conclusion} is the conclusion.

\section{Quantum Operators}
\label{sec:quantumoperators}

In this Section, we define the quantum operators used in this work. 
 Given a quantum state $\ket{\psi}$ containing qubits $a$ and $b$, we use
$CX_{(a,b)}\ket{\psi}$
to indicate a controlled-$X$ operation, where the control and the target are indicated by a subscript $a$ and $b$, respectively.

We define the one qubit operator $U^{(x_k,\gamma_k)}$ in Eq.~\ref{eq:matrixCU}. In CV-QRAM ~\cite{veras2020}, the value of  $\gamma_{k}=\gamma_{k-1} -|x_{k-1}|^2$ is a iteration variable,  with $1 \leqslant k \leqslant M-1$, and initial condition $\gamma_{0}=1$. Furthermore, $x_{k}=\alpha+i\beta$ is a complex number, with its complex conjugate denoted by $x_{k}^{*}$.

\begin{equation}
\label{eq:matrixCU}
U^{(x_k,\gamma_k)} = \left[\begin{array}{cc}
 \sqrt{\frac{\gamma_{k}-|x_{k}|^2}{\gamma_{k}}}  & \frac{x_{k}}{\sqrt{\gamma_{k}}} \\
 \frac{-x_{k}^{*}}{\sqrt{\gamma_{k}}} & \sqrt{\frac{\gamma_{k}-|x_{k}|^2}{\gamma_{k}}}\end{array}\right].
\end{equation}

Given a quantum state $\ket{\psi}$ with $n+1$ qubits, which contains the $t+1$ qubits $a_{0},a_{1},\ldots, a_{t-1},b$ with $t\leq n$, we use
${C^{t}U^{(x_k,\gamma_{k})}}_{(a_{0}a_1\ldots a_{t-1},b)}\ket{\psi}$
to denote a $t$-controlled operator, where the $t$ qubits $( a_0, a_1, \ldots, a_{t-1})$ are the controls of the operation, and $b$ denotes the target qubit, where we apply the $U^{(x_k,\gamma_{k})}$ operator if all the controls have a value $1$. In CV-QRAM, this controlled operation  is responsible for loading the complex number $ x_k $, as an amplitude associated with the  pattern $p_k$.

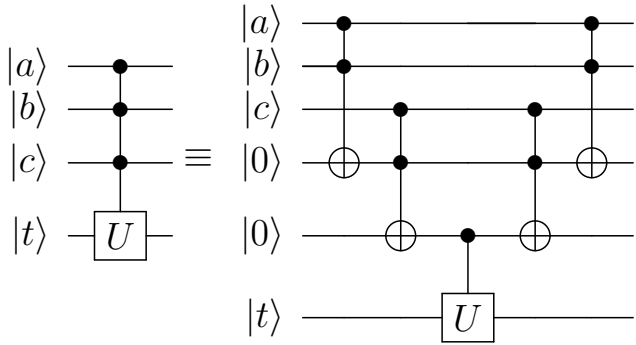
\begin{figure*}[ht]
\[\scalebox{1.2}{%
\Qcircuit @C=0.7 em @R=0.9em {
    &&&&&&&\lstick{\ket{a}}&\ctrl{1} &\qw&\qw&\qw\qw&\ctrl{1}&\qw \\
	\lstick{\ket{a}}&\ctrl{1}&\qw &&&&&\lstick{\ket{b}}& \ctrl{2}\qw&\qw&\qw&\qw&\ctrl{2}&\qw \\
	\lstick{\ket{b}}& \ctrl{1} & \qw &\raisebox{-6.5ex}{$\equiv$}&&&&\lstick{\ket{c}}&\qw&\ctrl{1}&\qw&\ctrl{1}&\qw&\qw\\
	\lstick{\ket{c}}& \ctrl{1}&\qw  &&&&&\lstick{\ket{0}}&\targ&\ctrl{1}&\qw&\ctrl{1}&\targ&\qw\\
	\lstick{\ket{t}}& \gate{U}&\qw &&&&&\lstick{\ket{0}}&\qw&\targ&\ctrl{1}&\targ&\qw&\qw\\
	\lstick{}& &&&&&&\lstick{\ket{t}}&\qw&\qw&\gate{U}&\qw&\qw&\qw\\
}} \]
\caption{Decomposition of a $C^{3}U$ as proposed in the \cite{nielsen2002quantum} }
\label{circuit-decomposition-C3U-Nielsen}
\end{figure*}
\

In current systems $n$-controlled gates $C^{n}U$ are not available as a native machine instruction. Since any quantum gate can be implemented using single-qubit  gates and the CNOT quantum operator \cite{barenco1995elementary}, we adopted a decomposition for an operator of the type $C^{n}U$ combining the decompositions in~\cite{nielsen2002quantum} and~\cite{barenco1995elementary}. According to~\cite{nielsen2002quantum} the quantum gate $C^{n}U$ can be decomposed into $2(n-1)$ Toffoli gates and one single quantum gate $CU$, using $n-1$ auxiliary qubits prepared in $\ket{0}$. For instance, Figure~\ref{circuit-decomposition-C3U-Nielsen} shows the decomposition of a $C^{3}U$.
Toffoli gates generated by this decomposition are applied in pairs. Then we can apply the decomposition proposed in \cite{barenco1995elementary}, where 3 CNOT gates are needed to implement a Toffoli gate. We can implement the $CU$ gate using 2 CNOT gates. Therefore, the decomposition of $C^nU$ operator uses $ 6n-4 $ CNOT gates.

\section{CV-QRAM Algorithm}\label{sec:CV-QRAM algorithm}

The CV-QRAM~\cite{veras2020} receives as input a data set (Eq.~\ref{eq:data}), and provides as output a quantum circuit $SP_\mathcal{D}$ with $SP_\mathcal{D}\ket{0}= \ket{\psi}$ (Eq.~\ref{eq:amp_enc}). CV-QRAM is a deterministic algorithm that performs an exact preparation of the desired quantum state, eliminates the post-selection of FF-QRAM~\cite{Park2019} and has computational cost $\mathcal{O}(Mn)$ steps to store $M$ input patterns with $n$ qubits.

The CV-QRAM algorithm is described in Algorithm~\ref{alg:CV-QRAM_pcode} and uses two quantum registers $\ket{u;m} $, where $\ket{u} =\ket{u_0u_1}$ is an auxiliary register initialized in $\ket{01}$, and $\ket{m}$ is a memory register initialized in $\ket{0}^{\otimes n}$.
 
 \begin{algorithm}
    \SetKwFunction{load}{cvqram}
    \SetKwInOut{Input}{input}\SetKwInOut{Output}{output}
    \Input{data = $\{x_k, p_k\}_{k=0}^{M-1}$}
    \BlankLine
    \Output{$\ket{m} = \sum_{k=0}^{M-1} x_k \ket{p_k}$}
    \BlankLine
    \Fn{\load($data$) \label{cv_step:1}}
    {
    	$\ket{\psi_{0_0}} = \ket{u}\ket{m} = \ket{01}\ket{0}^{\otimes n}$ \\ \label{cv_step:initial}
    	\ForEach{$(x_k,p_k) \in \data$}{ \label{cv_step:3}
    		\For{$j = 0 \to n-1$ \label{Apqm_step:4}}{
        		\uIf{$p_{k}[j] \neq 0$}
        		{
        		    $\ket{\psi_{k_1}} = CX_{u_1,m_j}\ket{\psi_{k_0}}$ \\ \label{rstep5Apqm_step:5}
        		}
        		\Else
        		{
        		    $\ket{\psi_{k_1}} = X_{m_j}\ket{\psi_{k_0}}$
        		} \label{Apqm_step:8}
    		}
    		$\ket{\psi_{k_2}} = C^{n}X_{m,u_0}\ket{\psi_{k_1}}$ \\ 
    		\label{Apqm_step:9}
    		$\ket{\psi_{k_3}} = {CU^{(x_k, \gamma_{k})}}_{u_0,u_1}\ket{\psi_{k_2}}$\\ 
    		\label{Apqm_step:10}
    		$\ket{\psi_{k_4}} = C^{n}X_{m,u_0}\ket{\psi_{k_3}}$ \\ 
    		\label{Apqm_step:11}
    		\For{$j = 0 \to n-1$\label{Apqm_step:12}}{
        		\uIf{$p_k[j] \neq 0$}
        		{
        		    $\ket{\psi_{k_5}} = CX_{u_1,m_j}\ket{\psi_{k_4}}$ \\ 
        		}
        		\Else
        		{
        		    $\ket{\psi_{k_5}} = X_{m_j}\ket{\psi_{k_4}}$\\ \label{Apqm_step:16}
        		}
    		}
    	}
    	\KwRet\ $\ket{m}$
	}
	\caption{CV-QRAM - Complex Data Storage Algorithm}
	\label{alg:CV-QRAM_pcode}
\end{algorithm}
 
\begin{figure*}[ht]
 \centering \includegraphics[width=0.8\textwidth]{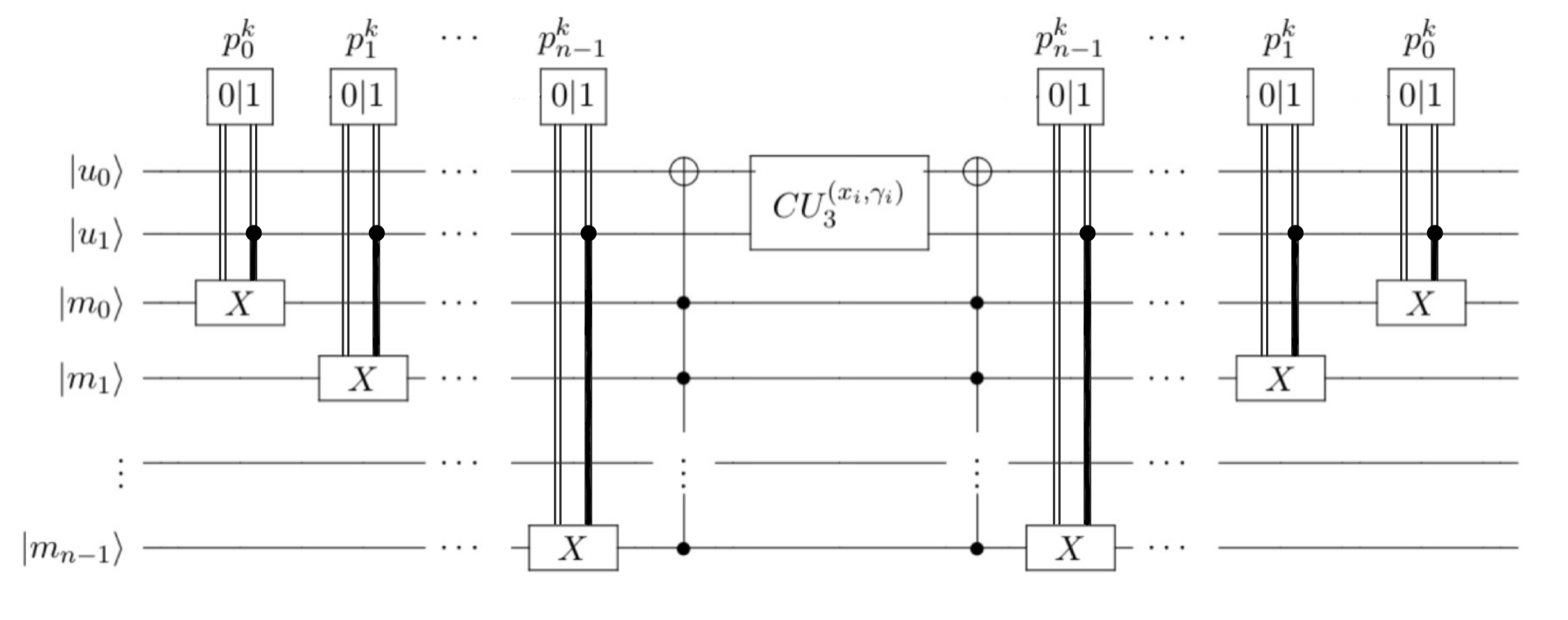}
\caption{One iteration of the CV-QRAM algorithm, to store a $(x_k,p_k)$ input pattern.}
\label{circ:CV-QRAM}
\end{figure*}

 In Algorithm~\ref{alg:CV-QRAM_pcode} the Step~\ref{cv_step:initial} initializes quantum registers $\ket{u}$ and $\ket{m}$. The for loop starting in Step~\ref{cv_step:3} is the main loop and loads the pattern $p_k$ with amplitude $x_k$ into $\ket{m}$. The for loop starting in Step~\ref{Apqm_step:4} converts $\ket{0}^{\otimes n}$ into $\ket{1}^{\otimes n}$ if $u_1=1$. Steps \ref{Apqm_step:9} to \ref{Apqm_step:11} initialize the amplitude $x_k$. The auxiliary qubit $ \ket{u_1} $ divides the quantum state into two parts: one containing the terms where the patterns are already stored, and the other containing the term that is being processed to receive a new pattern. The for loop starting in Step~\ref{Apqm_step:12} restores $\ket{1}^{\otimes n}$ to $\ket{0}^{\otimes n}$ if $u_1 = 1$ and changes $\ket{1}^{\otimes n}$ to $\ket{p_k}$ if $u_1 = 0$. Figure~\ref{circ:CV-QRAM} presents the quantum circuit corresponding to an iteration of CV-QRAM main loop for loading an input pattern $ (x_k, p_k) $ in $\ket{m}$.

\section{CV-QRAM Optimization Algorithm }\label{sec:CVO-QRAM}

The number of controlled operations required by the CV-QRAM algorithm depends on the number of input patterns $M$ in the data set $\mathcal{D}$ and the number  of qubits $n$. To store a pattern $(x_k,p_k)$ the cost of the CV-QRAM is always the same, regardless of the number of 1s in the patterns $p_k$.  
In CVO-QRAM, the controlled operations are determined by the number of bits with a value 1 in the patterns and the position where they occur. To achieve this reduction in the number of controlled operations we first define a partial order to the string of bits.
The patterns are ordered in relation to the number of 1s in the binary string. If $(x_r,p_r)$ is stored  in the quantum state before $(x_s, p_{s})$, it is because $||p_r||_2 \leqslant ||p_s||_2$, where $||p_k||_2$ denotes the Euclidean norm.

\subsection{ Defining controlled operations according to storage patterns}

PQM and CV-QRAM algorithms require at least one 
$C^{n}U$ operation for each iteration performed to load an input pattern. Let $t$ be the number of 1s in $p_k$, CVO-QRAM requires one $C^{t}U$ operation to load the input pattern $(x_k,p_k)$, with a total of $2t$ $CX$ operations and  one $t$-controlled operation  $C^{t}U$, arranged as follows:

\begin{itemize}
    \item[a)] One ${CX}_{(u,p_k[j])}$ operation for each $j$, where $p_k[j]=1$. 
    \item[b)] One operation $C^{t}U^{(x_k,\gamma_{k})}$ , with control in all $t$ qubits where $p_k[j]=1$ and target in $\ket{u}$.
    \item[c)] One ${CX}_{(u,p_k[j])}$ operation for each $j$, where $p_k[j]=1$, reversing the first ${CX}_{(u,p_k[j])}$ operations.
\end{itemize}

\begin{figure}[!htb]
\begin{subfigure}{0.6\textwidth}
  \includegraphics[width=\linewidth]{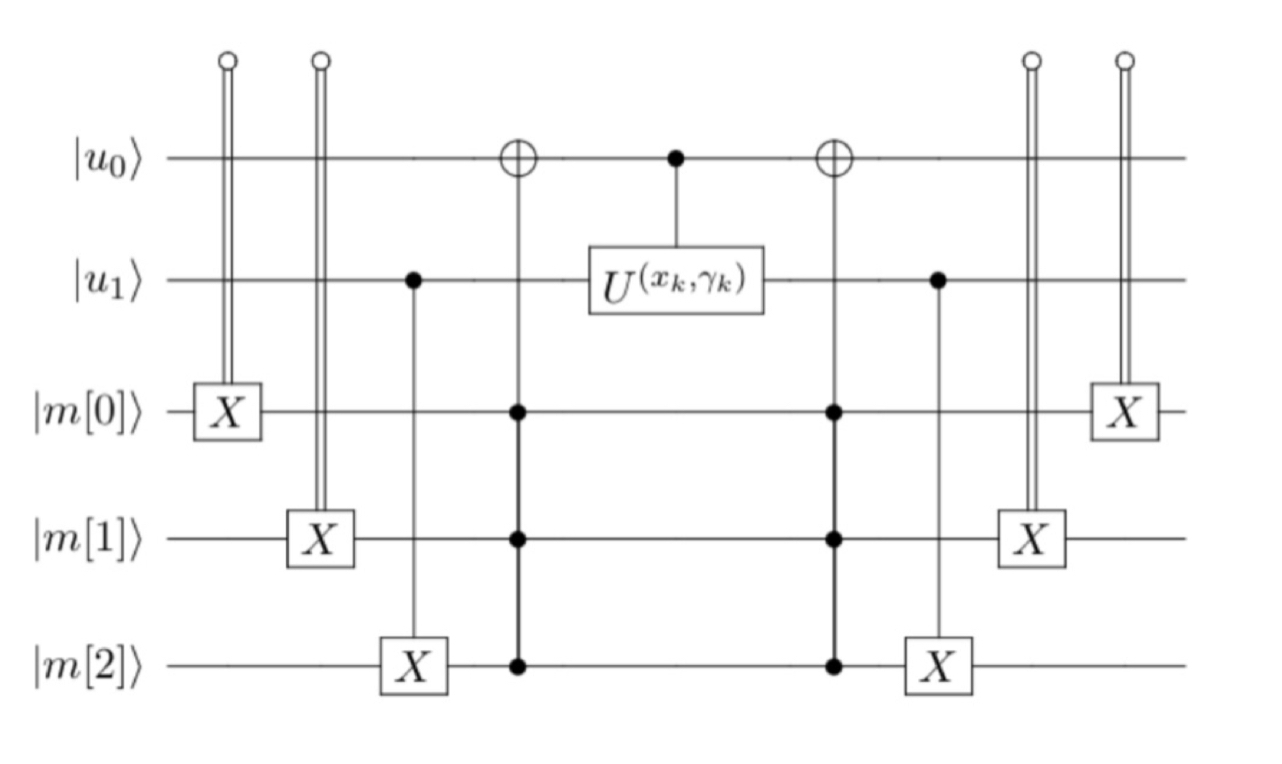}
  \caption{}\label{fig:pk_t=1}
\end{subfigure}\hfill
\begin{subfigure}{0.45\textwidth}
  \includegraphics[width=\linewidth]{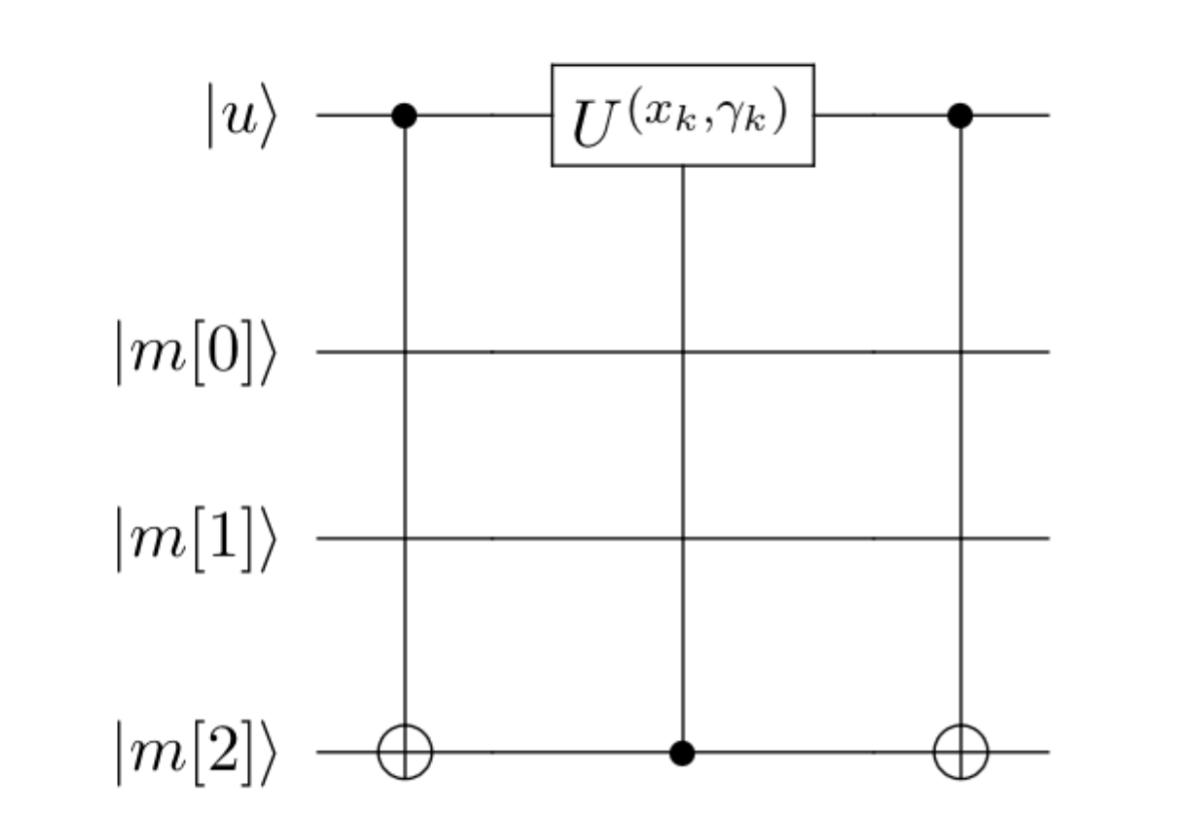}
  \caption{}\label{fig:pk_t=2}
\end{subfigure}
\caption{Loading $x_k|001\rangle$ with (a) CV-QRAM and (b) CVO-QRAM.}
\end{figure}

Fig.~\ref{fig:pk_t=1} and Fig.~\ref{fig:pk_t=2} show the circuits of the CV-QRAM and CVO-QRAM algorithms to store an input pattern, where $\ket{p_k} =\ket{001}$. The CV-QRAM algorithm needs 18 controlled operations. In contrast, CVO-QRAM needs 4 controlled operations, which shows a significant reduction in the number of CNOT gates when the binary string has few 1s. The CVO-QRAM cost is related with the number of patterns $p_k$, and the number of $ 1s $ of each $p_k\in \mathcal{D}$. 

The $U^{(x_k,\gamma_{k})}$ rotation operator is  responsible for loading the $ x_k $ amplitude in the $ p_k $ pattern, at the end of each algorithm iteration. Particularly, if all bits in the pattern are zero, then the $U$ operator without controls will be applied in the qubit $\ket{u}$. Another particular case happens when storing the last pattern. In this case, the last sequence of operations $CX$ does not need to be applied.

The auxiliary register $ \ket{u}$, works in a similar way to the auxiliary qubit $\ket{u_1}$ of CV-QRAM, it divides the quantum state into terms where the patterns are stored (in processing), indicated  by $\ket{u}=\ket {0}$ ($\ket{u}=\ket{1}$). At each iteration of the CVO-QRAM algorithm the $U^{(x_k,\gamma_{k})}$ operator responsible for loading a $ p_k $ pattern with associated amplitude $ x_k $ is applied to the  qubit $\ket{u}$. After the last iteration, the auxiliary qubit $\ket{u}$ will be in state $\ket{0}$.

Let $(x_k,p_k)$ be the input pattern that will be stored, $t$ the number of bits with value 1 in the binary string and  $l$ a list containing the positions where $p_k[j] =1$. If $t=0$ no controlled operation is necessary and we apply the  $U^{(x_k,\gamma_{k})}$ operator in the auxiliary qubit $\ket{u}$. If $t\geqslant 1$ according to the configuration of the $ p_k $ pattern, we have $2t+1$ controlled operations that will be performed in quantum state $\ket{\psi}$, as following:

    \begin{equation}\label{eq:pk_t=r}
    Q{C^{t}U}_{(m[l_0,l_1,\ldots l_{t-1}],u)} Q^\dagger \ket{\psi},
    \end{equation}
where $l_0, l_1, \dots, l_{t-1} \in l$  and $Q$ is the quantum operator, described below:

\begin{eqnarray*}
Q&=&CX_{(u,m[l_{t-1}])}  \cdots CX_{(u,m[l_1])} \cdot CX_{(u,m[l_0])}.\\
\end{eqnarray*}

\subsection{CVO-QRAM Algorithm}
\label{alg:optimization}

The CVO-QRAM algorithm is described in Algorithm~\ref{alg:CV-QRAM optimization}, and we suposse that patterns $p_k$ are ordered such that $\|p_k\|_2 \leq \|p_{k+1}\|_2$. The index $s$ in the state $\ket{\psi_{k_{s}}}$ denotes the step $s$ of the state preparation of the input pattern $(x_k,p_k)$.
\begin{algorithm}
    \SetKwFunction{load}{cvoqram}
    \SetKwInOut{Input}{input}\SetKwInOut{Output}{output}
    \Input{data = $\{x_k, p_k\}_{k=0}^{M-1}$}
    \BlankLine
    \Output{$\ket{m} = \sum_{k=0}^{M-1} x_k \ket{p_k}$}
    \BlankLine
    \Fn{\label{step:1 load}\load($data$, $\ket{\psi}$)}
    {
    	$\ket{\psi_{0_0}} = \ket{u}\ket{m} =
    	\ket{1;0,\cdots,0}$ \\ \label{step:2 initialstate CVO-QRAM}
    	\ForEach{$(x_k,p_k) \in \data$}{ \label{step:3 loop}
        	\label{rstep3}
    		\label{t} $t=$ The number of bits with value $ 1 $ in the pattern $ p_k $ \\
    	    \label{list} $l=$ a list containing the positions of $ p_k $ where $ p_k [j] =
    	    1.$ \\
    	    \BlankLine
    	    \BlankLine
    	  
    		$\ket{\psi_{k_1}} =\prod \limits_{l_i\in l} CX_{(u,m[l_i])}\ket{\psi_{k_0}}$ \\   \label{rstep6}
    		\BlankLine
    		\BlankLine
    		$\ket{\psi_{k_2}} = {{C^{t}U}_{(m[l_0,l_1,\ldots l_{t-1}],u)}}\ket{\psi_{k_1}}$\\\label{rstep7}
    		\BlankLine
    		\BlankLine
    		\If{$k \neq M-1$}
    		{$\ket{\psi_{k_3}} =\prod \limits_{l_i\in l} CX_{(u,m[_i])}\ket{\psi_{k_2}}$ \\\label{rstep8}}
    	}
    	\KwRet\ $\ket{m}$
	}
	\caption{CVO-QRAM - Complex Data Storage Algorithm}
	\label{alg:CV-QRAM optimization}
\end{algorithm}

 In Algorithm~\ref{alg:CV-QRAM optimization} the step \ref{step:2 initialstate CVO-QRAM} initializes the quantum registers $\ket{u}$ and $\ket{m}$, $\ket{u}=\ket{1}$ and $\ket{m}=\ket{0}^{\otimes n}$. The for loop starting in Step~\ref{step:3 loop} is the main loop. For each $(x_k, p_k) \in \mathcal{D}$ the variable $t$ denotes the number of bits with value $ 1 $ in the binary pattern $p_k$, while the variable $ l $ denotes a list containing the values $ j $, such that $ p_k [j] = 1 $ in $ p_k $. Step~\ref{rstep6} loads the binary string $p_k$ in the memory where the processing term $u$ is equal to one. Step~\ref{rstep7} initializes amplitude $x_k$. We only need $t$ controls because the stored patterns have at most $t$ 1s. Step~\ref{rstep8} restores memory in the processing term to $\ket{0}^{\otimes n}$, in the last iteration this step is not required because a superposition of states will not be created in the auxiliary qubit of the term in processing, since there will be no more input patterns to be loaded.

\subsection{CVO-QRAM Algorithm cost}

Most multiple qubit gates are not available in the instruction set of actual quantum devices. Implementing quantum gates that act simultaneously on two or more qubits is possible by decomposing them into gates that act on one qubit and CNOT gates~\cite{barenco1995elementary}. CNOT gates implementation are more susceptible to errors when compared to gates over one qubit \cite{Shende2006}. The amount of CNOT gates is one of the ways to evaluate the cost of state preparation algorithms \cite{mottonen2005transformation,Shende2006,plesch2011quantum}. We follow this strategy and use the number of CNOT gates to measure the efficiency of the CVO-QRAM algorithm.

We also consider the computational cost in the classical device. Given an input pattern $\mathcal{D}$ with $M$ patterns $(x_k,p_k)$ where $x_k$ are complex amplitudes and $p_k$ are binary patterns of $n$ bits. It is necessary to  sort the $M$ input patterns with computational cost $O(M\log M)$. For each pattern $p_k$ the algorithm adds $O(n)$ gates in the quantum circuit with a total of $O(nM)$ steps to create the circuit. The overall cost in the classical device is $O(M \log M + nM)$.

In general CVO-QRAM uses $O(nM)$ controlled gates. For dense quantum states ($M=2^n$) the number of CNOT gates is described in Eq.~\ref{CVO-cost-dense-f(n)}, where $C(n,t)$ is  the binomial coefficient. In this case, the CVO-QRAM needs more CNOT gates than state preparation algorithms for dense states.

\begin{equation}\label{CVO-cost-dense-f(n)}
  \sum_{t=1}^{n} C(n,t)(8t-4)-n,
\end{equation}

The advantage of CVO-QRAM is in the construction of sparse quantum states. The number of CNOT gates required to prepare a sparse quantum state with $ M \ll 2^n $ inputs is described in Eq.~\ref{cvo-sparse-cost}, where $\mu_t$ is the number of input patterns $ p_k $ in $ \mathcal{D} $ with $ t $ bits with value 1 in the binary string, and $t_{max}$ denotes the highest value of $t$.

\begin{equation}\label{cvo-sparse-cost}
\sum_{t=1}^{n} \mu_{t}(8t-4)-t_{max}, 
\end{equation}

We name double sparse states the states with $ 0< M\ll 2^n$ and $0\leqslant t_{max} \ll n$, this is the best case for the CVO-QRAM algorithm. The worst cases for the CVO-QRAM algorithm occur when the total number of bits with a value of 1 is close to $50\%$ . In both cases CVO-QRAM requires a smaller number of controlled operations than CV-QRAM.

\section{Results and Experiments}\label{sec:Results_and_experiments}

We compare the CNOT cost of the CVO-QRAM algorithm for state preparation presented in Section~\ref{sec:CVO-QRAM}  with other dense and sparse state preparation algorithms. All of the algorithms are implemented in \textit{Qiskit}~\cite{Qiskit}. 

\begin{table}[ht]
    \centering
    \begin{tabular}{l|l|l}
    Algorithm  & CNOT count &  Script  \\[1ex]
    \hline\hline \\[-1.5ex]
    CVO-QRAM (Section \ref{sec:CVO-QRAM})              & $\sum_{t=1}^{n} C(n,t)(8t-2)-n$ & \cite{adj2021} \\[1ex]
    CV-QRAM~\cite{veras2020}                           & $2^n (8n-2)$                    & \cite{adj2021} \\[1ex]
    UGD~\cite{plesch2011quantum}                       & $<\frac{23}{24}2^n$             & \cite{adj2021} \\[1ex]
    SQL~\cite{Shende2006}                              & $2^{n+1}-2n$                    & \cite{Qiskit} \\[1ex]
    Isometry~\cite{Iten2016}                           & $<\frac{23}{32}2^n$             & \cite{Qiskit} \\[1ex]
    M{\"o}tt{\"o}nen~\cite{mottonen2005transformation} & $ 2^{n+2}-4n-4$                 & \cite{Carstenblank21} \\[1ex]
    FF-QRAM~\cite{Park2019}                            & $2^n(6n-4)$                     & \cite{Carstenblank21} \\[1ex]
    \hline
    \end{tabular}
    \caption{ Theoretical number of CNOT gates and scripts used in the simulations to prepare a dense quantum state with $n$ qubits.}
    \label{tab:algorithms}
\end{table}

\subsection{Dense quantum state preparation}

\begin{figure}[ht]
\centering
\includegraphics[width=0.70\textwidth]{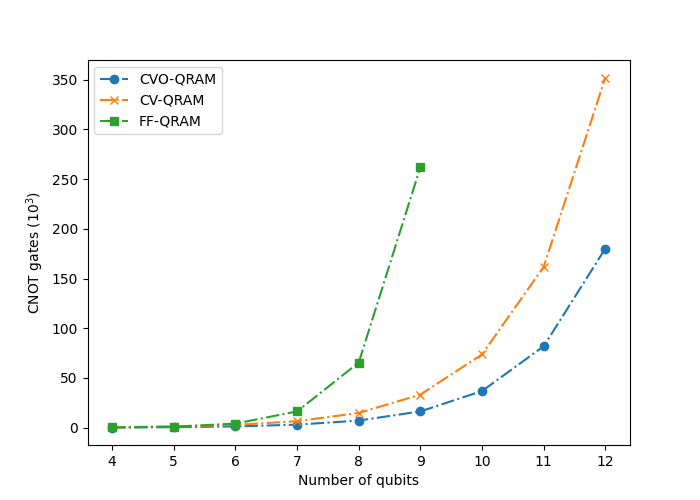}
\caption{CNOT count for prepare a dense quantum state with $n$ qubits.}\label{fig:dense-cvo-vs-cv}
\end{figure}

Table~\ref{tab:algorithms} summarizes the number of CNOT gates required by each quantum algorithm to prepare a dense quantum state with $n$ qubits and points to their public implementation. Although CVO-QRAM, CV-QRAM and FF-QRAM are not recommended in this case, we performed experiments with them to prepare a dense quantum state to compare these methods. The  results  are  presented  in  Figure~\ref{fig:dense-cvo-vs-cv}, FF-QRAM and CV-QRAM algorithms use significantly more CNOTs than the CVO-QRAM algorithm.

\subsection{Sparse quantum state preparation}

\begin{figure}[ht]
\centering
\includegraphics[width=0.8\textwidth]{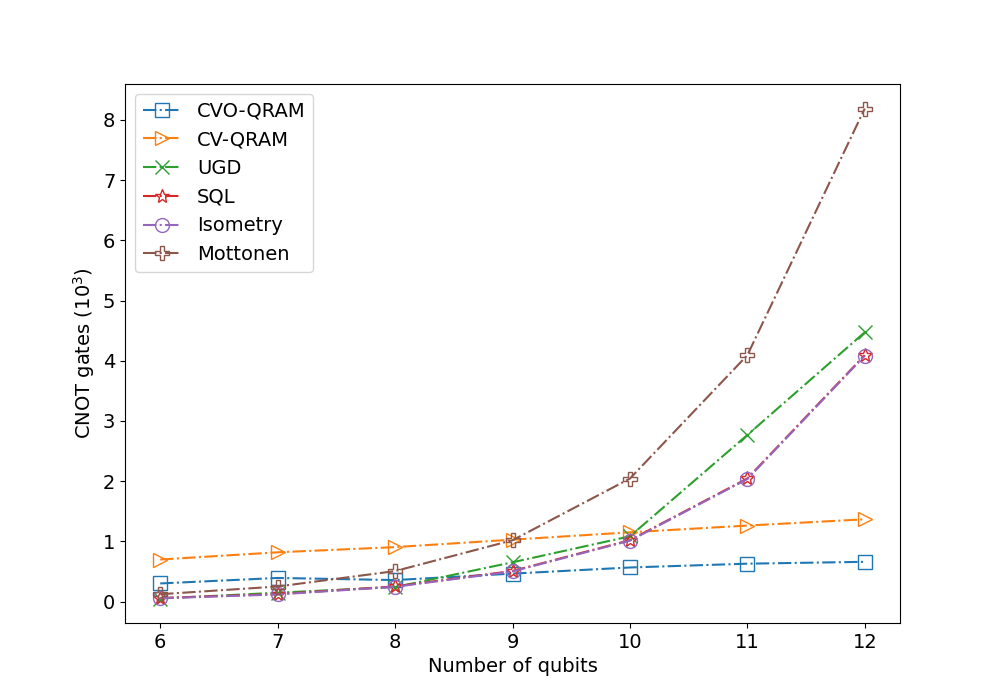}
\caption{Performance of sparse estate preparation. CNOT count for prepare a sparse quantum state of $n$ qubits with $M=2^4$ input data.}\label{fig:sparse-all}%
\end{figure} 

Figure~\ref{fig:sparse-all} presents the number of CNOTs gates needed to prepare a sparse quantum state with $n$ qubits and $M=2^4$ non-zero inputs. In this experiment, the set of binary patterns have about $50\%$ values equal to 1, this percentage of 1s in the binary string is the worst case for the CVO-QRAM algorithm. Note that as $n$ grows, the prepared quantum state becomes increasingly sparse and we can verify the CVO-QRAM advantage in the number of CNOTs.

\begin{figure}[ht]
    \begin{subfigure}{0.8\textwidth}
    \centering
    \includegraphics[width=\textwidth]{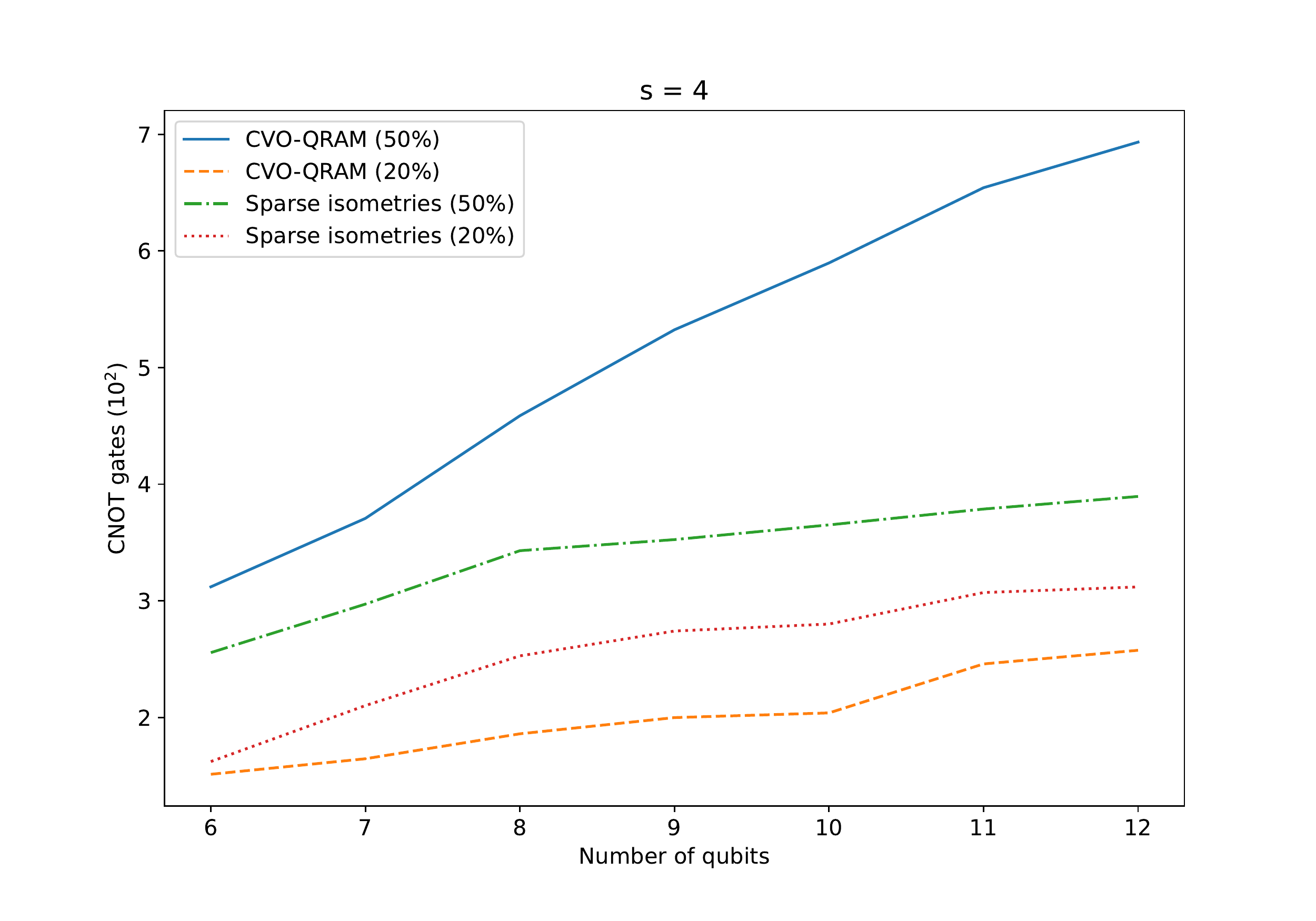}
    \caption{} \label{fig:pivotandcvo4}
    \end{subfigure} \newline
    \begin{subfigure}{0.8\textwidth}
    \centering
    \includegraphics[width=\textwidth]{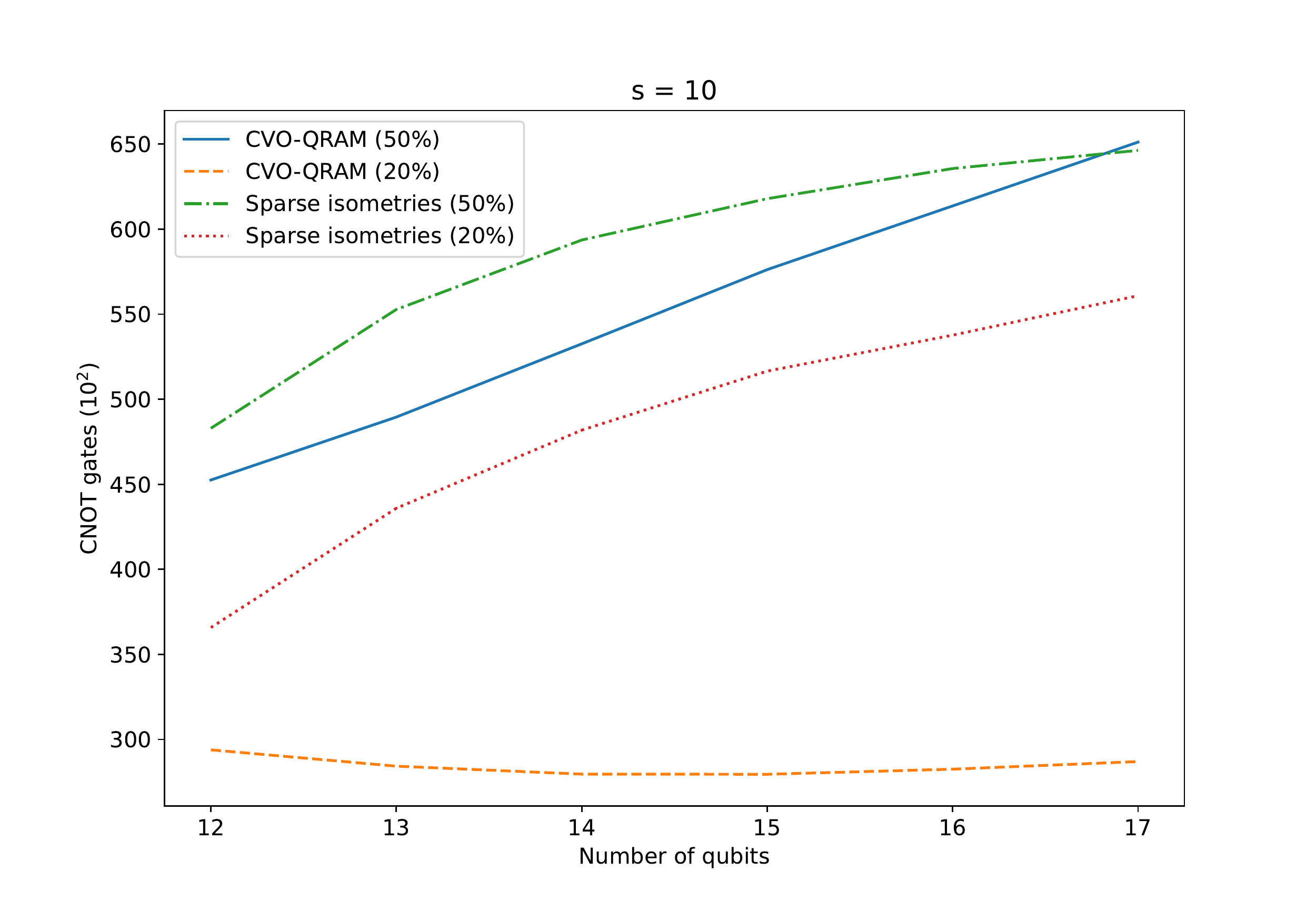}
    \caption{} \label{fig:pivotandcvo10}
    \end{subfigure}
    
    \caption{Average  number of CNOT gates produced by CVO-QRAM and the pivoting algorithm ~\cite{sparse_isometries_2021}  for  sparse  states  on $n$ qubits  with $2^s$ non-zero  entries. The 20\% and 50\% labels are approximately the average number of 1s in the binary strings $p_k$.  }    \label{fig:sparse-cvo-vs-cv}
\end{figure}

Recently an algorithm that deals with the preparation of sparse quantum states was proposed in \cite{sparse_isometries_2021}. The main idea of the sparse state preparation algorithm proposed in~\cite{sparse_isometries_2021} is to initialize a dense quantum state and then pivot the amplitudes to the correct position. Figure \ref{fig:sparse-cvo-vs-cv} shows simulations comparing the CVO-QRAM and the pivoting state preparation~\cite{sparse_isometries_2021} algorithms for sparse quantum states with $n$ qubits, containing $ 2^s $ non-zero amplitudes where $ s<n$. Both algorithms use the multi-controlled quantum gate decomposition described in Section~\ref{sec:quantumoperators} and are implemented in Qiskit \cite{Qiskit}.

In the double sparse case (label 20\%) the CVO-QRAM needs a smaller number of CNOTs than the pivoting algorithm for $s=4$ (Fig.~\ref{fig:pivotandcvo4}) and $s=10$ (Fig.~\ref{fig:pivotandcvo10}). In the CVO-QRAM worst case (label 50\%) the pivoting algorithm needs a smaller number of CNOTs for $s=4$ (Fig.~\ref{fig:pivotandcvo4}), but the CVO-QRAM is competitive for $s=10$ and $12\leq n<17$. Furthermore, the classical computational cost in \cite{sparse_isometries_2021} is quadratic with respect to the number of non-zero amplitudes compared to the CVO-QRAM which has time $O(nM)$.

\section{Conclusion}\label{sec:conclusion}

This work presented a quantum state preparation algorithm called CVO-QRAM that optimizes the CV-QRAM~\cite{veras2020} defining a partial order of the patterns to be stored. The CVO-QRAM is competitive with the state of art pivoting state preparation algorithm~\cite{sparse_isometries_2021} and requires a smaller number of CNOT gates in the double sparse case. The number of CNOT gates in CVO-QRAM is $8t-4$ CNOT gates per amplitude, where $t$ is the amount of 1s in the binary string of the pattern being stored. Thus CVO-QRAM depends on the number of input patterns  and the number of 1s in the binary strings.

To prepare an arbitrary quantum state is a task with exponential cost on the number of qubits. The sparse state preparation is an easier problem and has linear computational cost on the number of nonzero amplitudes. A possible future work is to identify classes of states that can be prepared more efficiently. For instance, the preparation of states with uniform amplitudes have been investigated in~\cite{mozafari2021efficient}. Another possible future work is to investigate the applicability of sparse state preparation in applications as solving linear systems of equations and quantum machine learning.

\section*{Data availability}
The site \url{https://www.cin.ufpe.br/~ajsilva/qclib} contains all the data and software generated during the current study.
\section*{Acknowledgments}
This work  is based upon research supported by CNPq, CAPES and FACEPE.

\end{document}